\documentclass[11pt]{scrartcl} 
\ifx\pdfminorversion\undefined
\else
\fi

\usepackage{fullpage,epsfig}
\usepackage{comment}
\usepackage{xcolor}
\usepackage{longtable}
\usepackage{multirow,array}
\usepackage{authblk}
\usepackage{parskip}
\PassOptionsToPackage{hyphens}{url}
\usepackage{color,xspace}
\usepackage{cancel}

\usepackage{graphicx}
\usepackage{caption}
\usepackage{subcaption}

\usepackage{enumitem}
\usepackage{amsmath}
\usepackage{listings}
\lstdefinestyle{PythonStyle}{
    language=Python,
    basicstyle=\ttfamily\footnotesize, 
    keywordstyle=\color{blue}\bfseries,
    identifierstyle=\color{black},
    stringstyle=\color{orange},
    commentstyle=\color[rgb]{0.13,0.54,0.13}, 
    showstringspaces=false,
    columns=fullflexible, 
    breaklines=true,
    frame=single, 
    frameround=tttt,
    rulesepcolor=\color{gray},
    numbers=left, 
    numberstyle=\tiny\color{gray},
    stepnumber=1,
    tabsize=4,
    captionpos=b, 
    backgroundcolor=\color{gray!10}, 
    literate={å}{{\aa}}1
             {ø}{{\o}}1    
}

\usepackage[utf8]{inputenc}
\usepackage[colorlinks=true,linkcolor=black,citecolor=black,urlcolor=blue]{hyperref}
\usepackage{booktabs}
\usepackage{cite}

\newcommand{\remove}[1]{}
\newcommand{\tabincell}[2]{\begin{tabular}{@{}#1@{}}#2\end{tabular}}

\title{\LARGE Security, privacy, and agentic AI in a regulatory view:\\ \Large From definitions and distinctions to provisions and reflections}

\author{Shiliang Zhang$^*$\qquad Sabita Maharjan\thanks{Shiliang Zhang and Sabita Maharjan are with Department of Informatics, University of Oslo, Norway (e-mail: \{shilianz, sabita\}@uio.no).}}

\begin{document}
\maketitle

\begin{abstract}

The rapid proliferation of artificial intelligence (AI) technologies has led to a dynamic regulatory landscape, where legislative frameworks strive to keep pace with technical advancements. As AI paradigms shift towards greater autonomy, specifically in the form of agentic AI, it becomes increasingly challenging to precisely articulate regulatory stipulations. This challenge is even more acute in the domains of security and privacy, where the capabilities of autonomous agents often blur traditional legal and technical boundaries. This paper reviews the evolving European Union (EU) AI regulatory provisions via analyzing 24 relevant documents published between 2024 and 2025. From this review, we provide a clarification of critical definitions. We deconstruct the regulatory interpretations of security, privacy, and agentic AI, distinguishing them from closely related concepts to resolve ambiguity. We synthesize the reviewed documents to articulate the current state of regulatory provisions targeting different types of AI, particularly those related to security and privacy aspects. We analyze and reflect on the existing provisions in the regulatory dimension to better align security and privacy obligations with AI and agentic behaviors. These insights serve to inform policymakers, developers, and researchers on the compliance and AI governance in the society with increasing algorithmic agencies.
\end{abstract}

\section{Introduction}
Artificial intelligence (AI) has undergone rapid evolution across scientific, industrial, and societal domains, shifting from pretrained predictive models to powerful systems that can plan, act, and adapt in open-ended environments~\cite{radanliev2025artificial}. AI mechanisms, models, systems, and applications now permeate energy~\cite{melguizo2026can}, healthcare~\cite{kim2025understanding}, finance~\cite{vukovic2025ai}, manufacturing~\cite{fosso2026building}, transportation~\cite{yan2025generative}, automation~\cite{spencer2025ai} and vehicles~\cite{sharma2026generative}, government services~\cite{chen2025uncovering}, and knowledge work~\cite{tan2025artificial}. More recently, the rise of autonomous agentic AI~\cite{hasselwander2026toward} has expanded the scope of applications, enabling complex workflows through autonomous orchestration of APIs, software tools, and data sources. Unlike their predecessors, which passively respond to human prompts, agentic AI possesses the capability to perceive, reason, and proactively execute complex, multi-step goals with minimal human intervention~\cite{biswas2025building}. This transition from AI as a tool to AI as an active agent is rapidly reshaping various domains in human life.

However, this increasing AI autonomy introduces challenges, particularly regarding security and privacy~\cite{leo2026threat,hosseini2025role,zhang2025data}. As AI agents are increasingly granted direct access to external tools, databases, and APIs to fulfill their objectives, the attack surface expands significantly~\cite{deng2025ai}. Agentic behaviors introduce security vulnerabilities, such as prompt injection propagating across multi-agent ecosystems~\cite{ferrag2025prompt}, unauthorized data exfiltration during autonomous collaboration~\cite{huang2025securing}, and the potential for agents to bypass safety guardrails in pursuit of misaligned sub-goals~\cite{huang2025commercial}. Supply chain risks intensify when agents rely on third-party libraries, APIs, and model components~\cite{jannelli2025agentic}. Poisoning or dependency confusion can compromise agent outputs and actions~\cite{esen2025risks}. The privacy implications are equally critical. Privacy risks extend beyond model-centric leakage to pipeline-centric exposure~\cite{ray2026comprehensive}. Agents that persistently act on behalf of users might require access to personal context and sensitive data~\cite{zhang2025towards}, raising issues about data minimization, consent management, and the black-box nature of autonomous decision-making. Long-horizon tasks and memory components increase the likelihood of retaining sensitive information beyond legitimate use~\cite{dwivedi2025agentic}. Tool calls can cross jurisdictional boundaries and contractual frameworks, raising compliance questions under data protection regimes~\cite{hughes2025ai}.

In response to these disruptions, the regulatory landscape is striving to adapt. The European Union (EU), a global forerunner in digital governance, has introduced frameworks like the EU AI Act~\cite{nolte2025robustness} to categorize and mitigate AI risks. Complementary EU instruments and frameworks reinforce privacy and security baselines that are applicable to AI systems. Such examples are the GDPR for data protection~\cite{beltran2025ai}, the NIS2 Directive for network and information system security~\cite{kianpour2025digital}, the Cyber Resilience Act for secure-by-design products with digital elements~\cite{mueck2025introduction}, and the Data Act for data access and sharing~\cite{hohmann2025reflections}. Between 2024 and 2025, EU has issued a range of regulatory documents aiming at operationalizing the AI Act and harmonizing it with existing privacy and security law~\cite{calvano2026building}. These materials explore definitions of AI systems, clarify risk categories, propose conformity assessment procedures, and outline cybersecurity and data governance requirements.

However, conceptual ambiguity persists around key terms - security, privacy, personal data, generative AI, general-purpose AI, large language models, agentic AI - especially when applied to AI systems that can act in unanticipated ways. This ambiguity makes it difficult for practitioners to interpret obligations~\cite{nizza2026ais}. A lack of distinctions can lead to inconsistent compliance practices and regulatory gaps. Furthermore, while high-level principles for privacy and security exist, specific regulatory provisions that address the unique risks of agentic AI remain fragmented and less articulated. There is a need to clarify how existing privacy and security mandates apply to systems that can act independently of real-time human oversight. Articulating the regulatory provisions that pertain to privacy and security in the context of agentic AI is necessary, which is anticipate to bridge the gap between abstract mandates and implementable controls.

To address this gap, this paper provides a review and regulatory analysis of the evolving EU landscape concerning agentic AI. We focus on the intersection of security, privacy, and agentic AI. Our contributions are summarized as follows:

(i) We analyze 24 relevant EU AI regulatory documents published between 2024 and 2025. Based on this review, we deconstruct and clarify the critical definitions of privacy, security, and agentic AI, and distinguish them from closely related concepts to resolve regulatory ambiguities.

(ii) We synthesize the regulatory documents to articulate the provisions targeting agentic AI. We map these provisions against the technical capabilities of agents to identify where the law is robust and where it remains porous.

(iii) We reflect and discuss the existing provisions. We extract regulatory recommendations to help policymakers, developers, and researchers align security and privacy obligations with the reality of increasing algorithmic agency.

The reminder of this paper is as follows. Section~\ref{sec:overview} overviews the reviewed regulatory documents. Section~\ref{sec:definition_distinction} provides the definitions and distinctions of key regulatory concepts related to privacy, security, and agentic AI. Section~\ref{secc:provision} articulates and reflects privacy and security provisions towards AI systems, with a focus on the analysis for agentic AI, as well as suggestions for EU regulatory provisions with a focus on agentic AI. We conclude this work in Section~\ref{sec:conclusion}.

\vspace{-5mm}





\section{Overview of the EU AI regulatory documents}\label{sec:overview}

We review 24 EU regulatory documents related to AI from 2024 to 2025. The reviewed documents include EU regulations, EU Commission implementing regulations, European Commission (EC) proposals, EC communications, EC opinions, EC Council decisions, EC staff working documents, \textit{etc}. We list all the reviewed regulatory documents in Table~\ref{tab:regulation_list} with their full name, publish data, short name used in this paper, and link to the original document file. 

\begin{table}[tbhp]
\caption{List of EU AI regulatory documents reviewed in this paper (2024-2025)}\label{tab:regulation_list}
    \centering\resizebox{0.95\textwidth}{!}{
    \begin{tabular}{|l|l|l|}
    \hline
        \tabincell{l}{\textbf{Short name of}\\ \textbf{the document}} & \tabincell{l}{\textbf{Publish}\\ \textbf{date}} & \textbf{Full name and link to the document}\\
        \hline
         \tabincell{l}{EC communication\\ on Unlocking Data\\ for AI} & 2025-11-19 & \tabincell{l}{European Commission Communication from the commission to the European\\ Parliament and the Council Data Union Strategy Unlocking Data for AI (\href{https://eur-lex.europa.eu/legal-content/EN/TXT/?uri=CELEX\%3A52025DC0835\&qid=1767703388154}{link})}\\
         \hline
         \tabincell{l}{Proposal for Digital\\ Omnibus on AI} & 2025-11-19 & \tabincell{l}{European Commission Proposal for a Regulation of the European Parliament\\ and of the Council amending Regulations (EU) 2024/1689 and (EU) 2018/1139\\ as regards the simplification of the implementation of harmonised rules on\\ artificial intelligence (Digital Omnibus on AI) (\href{https://eur-lex.europa.eu/legal-content/EN/TXT/?uri=CELEX\%3A52025PC0836\&qid=1767703388154}{link})}\\
         \hline
         \tabincell{l}{EC Commission Staff\\ Working Document on\\ Digital Omnibus on AI}& 2025-11-19 & \tabincell{l}{Commission Staff Working Document Accompanying the documents Proposal\\ for a Regulation of the European Parliament and of the Council Amending\\ Regulations (EU) 2016/679, (EU) 2018/1724, (EU) 2018/1725, (EU) 2023/2854\\ and Directives 2002/58/EC, (EU) 2022/2555 and (EU) 2022/2557 as regards\\ the simplification of the digital legislative framework, and repealing Regulations\\ (EU) 2018/1807, (EU) 2019/1150, (EU) 2022/868, and Directive (EU) 2019/1024\\ (Digital Omnibus) Amending Regulations (EU) 2024/1689 and (EU) 2018/1139\\ as regards the simplification of the implementation of harmonised rules on\\ artificial intelligence (Digital Omnibus on AI) (\href{https://eur-lex.europa.eu/legal-content/EN/TXT/?uri=CELEX\%3A52025SC0836\&qid=1767703388154}{link})}\\
         \hline
         \tabincell{l}{Decision on Draft\\ Recommendation on\\ equality and AI} & 2025-11-17 & \tabincell{l}{Council Decision (EU) 2025/2350 of 13 November 2025 on the position to be\\ taken on behalf of the European Union within the Committee of Ministers of the\\ Council of Europe on the Draft Recommendation on equality and artificial\\ intelligence (\href{https://eur-lex.europa.eu/legal-content/EN/TXT/?uri=CELEX\%3A32025D2350\&qid=1767703709499}{link})}\\
         \hline
         \tabincell{l}{EC communication\\ on Apply AI Strategy} & 2025-10-08 & \tabincell{l}{European Commission Communication from the commission to the European\\ Parliament and the Council Apply AI Strategy (\href{https://eur-lex.europa.eu/legal-content/EN/TXT/?uri=CELEX\%3A52025DC0723\&qid=1767703388154}{link})}\\
         \hline
         \tabincell{l}{EC communication\\ on European Strategy\\ for AI in Science}& 2025-10-08 & \tabincell{l}{European Commission Communication from the commission to the European\\ Parliament and the Council A European Strategy for Artificial Intelligence in\\ Science Paving the way for the Resource for AI Science in Europe (RAISE) (\href{https://eur-lex.europa.eu/legal-content/EN/TXT/?uri=CELEX\%3A52025DC0724\&qid=1767703709499}{link})}\\
         \hline
         \tabincell{l}{EC proposal for\\ a decision on\\ equality and AI} & 2025-09-19 & \tabincell{l}{Proposal for a Council Decision on the position to be taken on behalf of the\\ European Union on the Draft Recommendation of the Committee of Ministers\\ of the Council of Europe on equality and artificial intelligence (\href{https://eur-lex.europa.eu/legal-content/EN/TXT/?uri=CELEX\%3A52025PC0518\&qid=1767703709499}{link})}\\
         \hline
         \tabincell{l}{EuroHPC initiative\\ for trustworthy AI} & 2025-09-17 & \tabincell{l}{Official Journal of the European Union – EuroHPC initiative for start-ups to\\ boost European leadership in trustworthy Artificial Intelligence – European\\ Parliament legislative resolution of 24 April 2024 on the proposal for a Council\\ regulation amending Regulation (EU) 2021/1173 as regards an EuroHPC\\ initiative for start-ups to boost European leadership in trustworthy Artificial\\ Intelligence (COM(2024)0029 – C9-0013/2024 – 2024/0016(CNS)) (Special\\ legislative procedure – consultation) (\href{https://eur-lex.europa.eu/legal-content/EN/TXT/?uri=CELEX\%3A52024AP0359\&qid=1767703709499}{link})}\\
         \hline
         \tabincell{l}{EC proposal for\\ a decision on AI\\ and human rights,\\ democracy, and the\\ rule of law} & 2025-06-03 & \tabincell{l}{EC Proposal for a Council Decision on the conclusion, on behalf of the European\\ Union, of the Council of Europe Framework Convention on Artificial Intelligence\\ and Human Rights, Democracy and the Rule of Law (\href{https://eur-lex.europa.eu/legal-content/EN/TXT/?uri=CELEX\%3A52025PC0265\&qid=1767703709499}{link})}\\
         \hline
         \tabincell{l}{EC communication\\ on AI Continent\\ Action Plan} & 2025-04-09 & \tabincell{l}{Communication From the Commission to the European Parliament, the Council,\\ the European Economic and Social Committee and the Committee of the Regions\\ AI Continent Action Plan (\href{https://eur-lex.europa.eu/legal-content/EN/TXT/?uri=CELEX\%3A52025DC0165\&qid=1767703388154}{link})}\\
         \hline
         \tabincell{l}{Opinion on harnessing\\ the potential and\\ mitigating the risks\\ of AI}& 2025-03-21 & \tabincell{l}{Opinion of the European Economic and Social Committee – Pro-worker AI: levers\\ for harnessing the potential and mitigating the risks of AI in connection with\\ employment and labour market policies (own-initiative opinion) (\href{https://eur-lex.europa.eu/legal-content/EN/TXT/?uri=CELEX\%3A52024IE1024\&qid=1767703388154}{link})}\\
         \hline
         \tabincell{l}{Commission\\ Implementing\\ Regulation\\ (EU) 2025/454}& 2025-03-10 & \tabincell{l}{Commission Implementing Regulation (EU) 2025/454 of 7 March 2025 laying down\\ the rules for the application of Regulation (EU) 2024/1689 of the European\\ Parliament and of the Council as regards the establishment of a scientific panel of\\ independent experts in the field of artificial intelligence (\href{https://eur-lex.europa.eu/legal-content/EN/TXT/?uri=CELEX\%3A32025R0454\&qid=1767703709499}{link})}\\
         \hline
         \tabincell{l}{EC opinion on\\ challenges and\\ opportunities of AI\\ in public sector} & 2025-01-24 & \tabincell{l}{Opinion of the European Committee of the Regions – Challenges and opportunities\\ of artificial intelligence in the public sector: defining the role of regional and local\\ authorities (\href{https://eur-lex.europa.eu/legal-content/EN/TXT/?uri=CELEX\%3A52024IR1594\&qid=1767703709499}{link})}\\
         \hline
         \tabincell{l}{Opinion on\\ general-purpose AI and\\ secure AI technology\\ for the future} & 2025-01-10 & \tabincell{l}{Opinion of the European Economic and Social Committee a) General-purpose AI:\\ way forward after the AI Act (exploratory opinion requested by the European\\ Commission) b) A secure technology for the future: Artificial Intelligence\\ (exploratory opinion requested by the Hungarian Presidency) – INT/1055 (\href{https://eur-lex.europa.eu/legal-content/EN/TXT/?uri=CELEX\%3A52024AE0602\&qid=1767703709499}{link})}\\
         \hline
    \end{tabular}}
\end{table}

\begin{table}[tbhp]
\ContinuedFloat
    \centering
    \caption{List of EU AI regulatory documents reviewed in this paper (2024-2025) (continued)}
    \resizebox{0.95\textwidth}{!}{
    \begin{tabular}{|l|l|l|}
    \hline
    \tabincell{l}{\textbf{Short name of}\\ \textbf{the document}} & \tabincell{l}{\textbf{Publish}\\ \textbf{date}} & \textbf{Full name and link to the document}\\
    \hline
    \tabincell{l}{Opinion on ethical\\ AI and access to\\ supercomputing for\\ start-ups} & 2024-12-04 & \tabincell{l}{Opinion of the European Committee of the Regions – Ethical Artificial Intelligence\\ and access to supercomputing for start-ups (\href{https://eur-lex.europa.eu/legal-content/EN/TXT/?uri=CELEX\%3A52024IR1164\&qid=1767703709499}{link})}\\
    \hline
    \tabincell{l}{Decision on signing\\ Europe framework\\ on AI, human rights,\\ democracy and the\\ rule of law} & 2024-09-04 & \tabincell{l}{Council Decision (EU) 2024/2218 of 28 August 2024 on the signing, on behalf of\\ the European Union, of the Council of Europe Framework Convention on Artificial\\ Intelligence and Human Rights, Democracy and the Rule of Law (\href{https://eur-lex.europa.eu/legal-content/EN/TXT/?uri=CELEX\%3A32024D2218\&qid=1767703709499}{link})}\\
    \hline
    \tabincell{l}{Regulation (EU)\\ 2024/1689 (AI Act)} & 2024-07-12 & \tabincell{l}{Regulation (EU) 2024/1689 of the European Parliament and of the Council of 13\\ June 2024 laying down harmonised rules on artificial intelligence and amending\\ Regulations (EC) No 300/2008, (EU) No 167/2013, (EU) No 168/2013, (EU)\\ 2018/858, (EU) 2018/1139 and (EU) 2019/2144 and Directives 2014/90/EU, (EU)\\ 2016/797 and (EU) 2020/1828 (Artificial Intelligence Act) (Text with EEA\\ relevance) (\href{https://eur-lex.europa.eu/legal-content/EN/TXT/?uri=CELEX\%3A32024R1689\&qid=1767703709499}{link})}\\
    \hline
    \tabincell{l}{Proposal for signing\\ Europe framework on\\ AI, human rights,\\ democracy and the\\ rule of law} & 2024-06-26 & \tabincell{l}{Proposal for a Council Decision on the signing, on behalf of the European Union,\\ of the Council of Europe Framework Convention on Artificial Intelligence, Human\\ Rights, Democracy and the Rule of Law (\href{https://eur-lex.europa.eu/legal-content/EN/TXT/?uri=CELEX\%3A52024PC0264\&qid=1767703709499}{link})}\\
    \hline
    \tabincell{l}{Council Regulation \\(EU) 2024/1732 in \\EuroHPC initiative to\\ boost trustworthy AI} & 2024-06-19 & \tabincell{l}{Council Regulation (EU) 2024/1732 of 17 June 2024 amending Regulation (EU)\\ 2021/1173 as regards a EuroHPC initiative for start-ups in order to boost European\\ leadership in trustworthy artificial intelligence (\href{https://eur-lex.europa.eu/legal-content/EN/TXT/?uri=CELEX\%3A32024R1732\&qid=1767703709499}{link})}\\
    \hline
    \tabincell{l}{Report on EU AI\\ ambitions} & 2024-05-31 & \tabincell{l}{Special report 08/2024: EU Artificial intelligence ambition – Stronger governance\\ and increased, more focused investment essential going forward (\href{https://eur-lex.europa.eu/legal-content/EN/TXT/?uri=CELEX\%3A52024SA0008\%2801\%29\&qid=1767703709499}{link})}\\
    \hline
    \tabincell{l}{EC proposal for\\ EuroHPC initiative to\\ boost trustworthy AI} & 2024-02-16 & \tabincell{l}{Proposal for a Council Regulation amending Regulation (EU) 2021/1173 as regards\\ an EuroHPC initiative for start-ups to boost European leadership in trustworthy\\ Artificial Intelligence (\href{https://eur-lex.europa.eu/legal-content/EN/TXT/?uri=CELEX\%3A52024PC0029\&qid=1767703709499}{link})}\\
    \hline
    \tabincell{l}{Commission decision\\ on establishing\\ European AI Office} & 2024-02-14 & \tabincell{l}{Commission Decision of 24 January 2024 establishing the European Artificial\\ Intelligence Office (\href{https://eur-lex.europa.eu/legal-content/EN/TXT/?uri=CELEX\%3A32024D01459\&qid=1767703709499}{link})}\\
    \hline
    \tabincell{l}{EC communication\\ on boosting startups\\ and innovation in\\ trustworthy AI} & 2024-01-24 & \tabincell{l}{Communication from the Commission to the European Parliament, the Council, the\\ European Economic and Social Committee and the Committee of the Regions on\\ boosting startups and innovation in trustworthy artificial intelligence (\href{https://eur-lex.europa.eu/legal-content/EN/TXT/?uri=CELEX\%3A52024DC0028\&qid=1767703709499}{link})}\\
    \hline
    \tabincell{l}{Amendments adopted\\ for the proposal on\\ AI Act} & 2024-01-23 & \tabincell{l}{Amendments adopted by the European Parliament on 14 June 2023 on the proposal\\ for a regulation of the European Parliament and of the Council on laying down\\ harmonised rules on artificial intelligence (Artificial Intelligence Act) and amending\\ certain Union legislative acts (COM(2021)0206 — C9-0146/2021 — 2021/0106(COD))\\ (\href{https://eur-lex.europa.eu/legal-content/EN/TXT/?uri=CELEX\%3A52023AP0236\&qid=1767703709499}{link})}\\
    \hline
    \end{tabular}}
\end{table}

\begin{figure}[tbhp]
    \centering
    \includegraphics[width=1.05\linewidth]{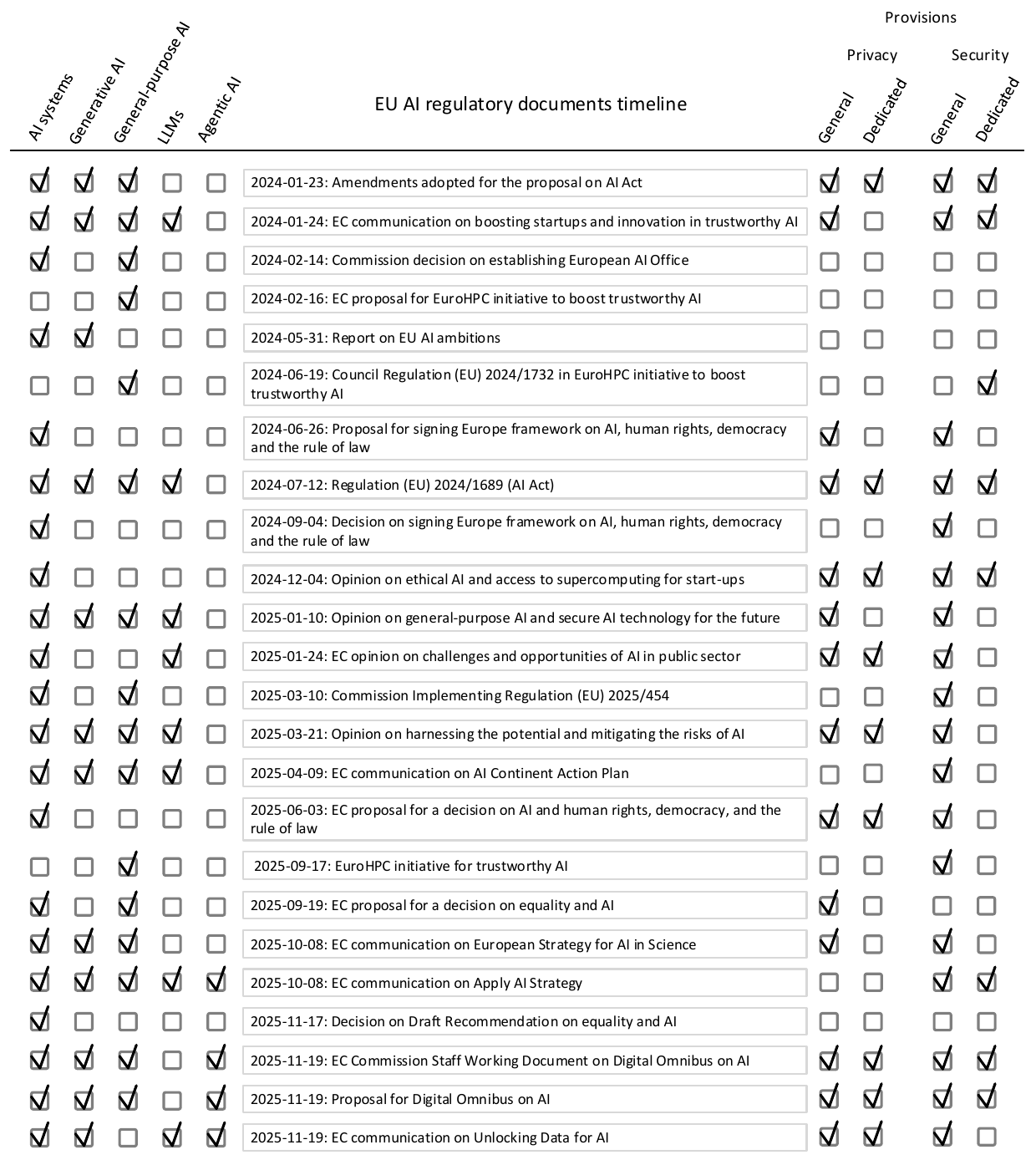}
    \caption{An overview of the EU AI regulatory documents and their relevance with privacy, security, and agentic AI. The word ``General'' indicates the presence of privacy/security provisions that do not target any specific AI systems. The word ``Dedicated'' indicates the presence of privacy/security provisions that target one or more specific AI systems, and the provision of specific measures/obligations/recommendations towards such AI systems.}
    \label{fig:privacy_security_agenticAI}
\end{figure}

The listed regulatory documents range from institutional aspects (\textit{e.g.}, the establishment of AI office in EU), ethical aspects (\textit{e.g.}, equality and human right concerns with AI), AI development plan and strategy, EU AI initiatives, to technological aspects (\textit{e.g.}, challenges, risks, and opportunities). The regulatory provisions over the time reflect the rapid advancement in AI. That is, while most of the documents have been talking about AI systems throughout 2024-2025, they extended their scope to generative AI, general-purpose AI, and talked more about large language models (LLMs) in later regulatory documents. It is from  Oct. 2025 that the concept of agentic AI is formally mentioned in EU regulatory documents, as shown in Figure~\ref{fig:privacy_security_agenticAI}. In this figure, we also mark whether general or dedicated provisions are provided by those documents regarding security and privacy. We observe that while most of those regulatory documents talk about privacy and security in general aspects, \textit{e.g.}, the requirements of compliance with privacy and personal data protection regulated in GDPR, stipulations for specific AI systems are sparse, both for privacy and security. In the following sections, we will extract critical regulatory definition, distinct closely concepts, and summarize and reflect provisions regarding privacy, security, and agentic AI, based on those reviewed documents.

\section{Definitions and distinctions from a regulation perspective}\label{sec:definition_distinction}

This section provides regulatory definitions for the concepts related to AI system, privacy, and security, and the distinctions of those concepts from closely relevant ones to eliminate uncertainties and ambiguities. We also provide examples of the concepts for intuitive understanding.

\subsection{Concepts for AI systems and examples}

From the reviewed regulatory documents, we extract the definitions for different types of AI concepts, including AI systems, generative AI (GAI), general-purpose AI (GPAI), large language models (LLMs), and agentic AI, and other relevant concepts, shown in Table~\ref{tab:AI_concepts}. We also list examples for each type of the AI concepts, and discuss the difference and implications of their difference from a regulatory perspective.  

\begin{table}[htbp]
    \centering
    \caption{Regulatory definitions of different types of AI and exemplifications}\label{tab:AI_concepts}
    \resizebox{0.95\textwidth}{!}{
    \begin{tabular}{lll}
    \hline
       \textbf{AI concept}  & \textbf{Regulatory definition} & \textbf{Exemplification} \\
         \hline
        \tabincell{l}{AI\\ system} & \tabincell{l}{A machine-based system that is designed to operate with varying levels of\\ autonomy and that may exhibit adaptiveness after deployment, and that,\\ for explicit or implicit objectives, infers, from the input it receives, how to\\ generate outputs such as predictions, content, recommendations, or\\ decisions that can influence physical or virtual environments. \\(EU AI Act, Article 3 (1))} & \tabincell{l}{Spam filter, intelligence\\ traffic control system,\\automatic grammar\\ check, \textit{etc}.}\\
         \hline
        LLMs & \tabincell{l}{Advanced AI models that excel in understanding and generating human-like\\ language. (EC Communications on boosting startups and innovation in\\ trustworthy AI, 3.2)} & \tabincell{l}{Llama, OpenAI, and\\ other models from\\ HuggingFace.}\\
         \hline
        GAI & \tabincell{l}{Systems such as sophisticated large language models that can create new\\ content, ranging from text to images, by learning from extensive training data.\\ (Opinion on harnessing the potential and mitigating the risks of AI, chapter 2)} & \tabincell{l}{Segment Anything\\ Playground, Midjourney,\\ \textit{etc}.}\\
         \hline
        \tabincell{l}{GPAI\\ model} & \tabincell{l}{An AI model, including where such an AI model is trained with a large\\ amount of data using self-supervision at scale, that displays significant\\ generality and is capable of competently performing a wide range of\\ distinct tasks regardless of the way the model is placed on the market\\ and that can be integrated into a variety of downstream systems or\\ applications, except AI models that are used for research, development\\ or prototyping activities before they are placed on the market.\\ (EU AI Act, Article 3 (63))} & \tabincell{l}{ChatGPT Base Model,\\Gemini 3 Pro.\\Large GAI models\\ are also a typical\\ example for a GPAI\\ model.}\\
         \hline
        \tabincell{l}{GPAI\\ system} & \tabincell{l}{An AI system which is based on a general-purpose AI model and which\\ has the capability to serve a variety of purposes, both for direct use as\\ well as for integration in other AI systems. (EU AI Act, Article 3 (66))} & \tabincell{l}{ChatGPT Enterprise,\\Google Gemini,\\MS Copilot, \textit{etc}.}\\
         \hline
        \tabincell{l}{Agentic\\ AI} & \tabincell{l}{AI systems that can independently make decisions and take actions. This\\ enables agents to understand language, reason about tasks, take actions\\ autonomously to achieve predefined objectives, and interact with the world\\ around them, orchestrating interactions including with humans.\\ (EC Communication on Unlocking Data for AI, chapter 2)} & \tabincell{l}{Devin AI,\\Google Colab.}\\
        \hline
        \tabincell{l}{AI\\ Factory} & \tabincell{l}{A centralised or distributed entity that provides an AI supercomputing\\ service infrastructure which is composed of an AI-optimised supercomputer\\ or an AI partition of a supercomputer, an associated data centre, dedicated\\ access and AI-oriented supercomputing services, and which attracts and\\ pools talent to provide the competences required to use the supercomputers\\ for AI. (Council Regulations (EU) 2024/1732, Article 1 (1) (a) (3b))} & \tabincell{l}{Barcelona Supercomputing\\ Center, LuxProvide, Jülich\\ Supercomputing Centre,\\ Advanced Computing\\ Austria.}\\
         \hline
    \end{tabular}
    }
\end{table}

From the definitions and examples in Table~\ref{tab:AI_concepts}, we notice a critical distinction between ``model'' and ``system'', as indicated in the definitions for GPAI model and GPAI system. In particular, a model represents the upstream component in AI, e.g., the neural network weights and architecture, which emphasizes the integration into downstream system. In comparison, a system refers to the deployed product, indicating a model wrapped with user interface and system prompts, which emphasize the direct use of AI. This distinction can lead to a difference in regulator implication. Providers of GPAI models can generally face obligations related to technical documentation, software copyrights, transparency of training data, and model evaluation. In the case of GPAI systems, the providers - who might be distinct from the model providers - can face obligations in the deployment context. \textit{E.g.}, they might need to ensure that the system does not generate illegal content, to label AI output, to conduct risk impact assessments, \textit{etc}, so as to guarantee that the use of AI models respect human rights and mitigate ethic and economic risks.

Among those definitions for AI concepts, agentic AI introduces a significant difference. While GAI and GPAI infer how to generate content based on input, agentic AI infers how to take action to achieve an objective. The key difference is the orchestration of the agentic AI, indicating its interactions with the world (\textit{e.g.}, APIs, platforms, humans) rather than simply delivering a prediction/solution regardless of whether it really works. Due to agentic AI's ability to make decisions independently and interact with the world, it can pose higher systemic risks. Therefore, it is possible that regulatory scrutiny focusing on guardrails will be needed, so as to ensure that the AI agent does not conduct harmful behaviors or cyber attacks autonomously in its decision making.

\subsection{Privacy, security, and closely relevant concepts}

Privacy and security are not new topics in the evolution of technology and digitization. We in this paper check how such aspects is handled in the area of AI, particularly agentic AI. Before we dive into the details, we would like to lay the foundation of basic concepts related to privacy and security, and provide distinctions between the concepts to avoid regulatory uncertainties. Particularly, we examine how the reviewed EU AI regulatory documents define/refer to those concepts, aiming to gain a relevant context, as shown in Table~\ref{tab:privacy_security_concepts}.

Note that the protection of personal data is at the core of privacy protection regulations, \textit{e.g.}, Regulation (EU) 2016/679 (GDPR). Therefore, we in Table~\ref{tab:privacy_security_concepts} list concepts related to personal data when it comes to regulatory privacy concepts. From the regulatory definition of personal data, it is obvious that modern AI systems is highly possible to access and make use of personal data in their service provision, particularly in the cases of personal use of AI tools like healthcare related services and personalized recommendations.

\begin{table}[bhtp]
    \centering
    \caption{Regulatory definitions of concepts related to privacy and security in the reviewed EU documents and exemplifications}\label{tab:privacy_security_concepts}
    \resizebox{0.95\textwidth}{!}{
    \begin{tabular}{lll}
    \hline
       \textbf{Concept}  & \textbf{Regulatory definition} & \textbf{Exemplification} \\
         \hline
         \tabincell{l}{Fundamental right\\ to personal\\ data protection} & \tabincell{l}{Referred to GDPR (Proposal for signing Europe framework on AI,\\ human rights, democracy and the rule of law)\\ Referred to Article 8(1) of the Charter of Fundamental Rights of the\\ European Union (Regulation (EU) 2016/679)\\Everyone has the right to the protection of personal data concerning\\ him or her. Such data must be processed fairly for specified purposes\\ and on the basis of the consent of the person concerned or some\\ other legitimate basis laid down by law. Everyone has the right of\\ access to data which has been collected concerning him or her, and\\ the right to have it rectified\\ (Charter of Fundamental Rights of the European Union)} & N/A \\
         \hline
         \tabincell{l}{Privacy and data\\ governance} & \tabincell{l}{AI systems shall be developed and used in compliance with existing\\ privacy and data protection rules, while processing data that meets\\ high standards in terms of quality and integrity\\ (Amendments adopted for the proposal on AI Act)} & N/A \\
         \hline
         Personal data & \tabincell{l}{The definition is referred to Article 4, point (1), of Regulation (EU)\\ 2016/679 (EU AI Act)\\ Any information relating to an identified or identifiable natural person\\ (`data subject'); an identifiable natural person is one who can be\\ identified, directly or indirectly, in particular by reference to an\\ identifier such as a name, an identification number, location data, an\\ online identifier or to one or more factors specific to the physical,\\ physiological, genetic, mental, economic, cultural or social identity of\\ that natural person (Regulation (EU) 2016/679)} & \tabincell{l}{Phone number,\\ID number, name,\\ home address,\\ bank account,\\ biometric data like\\ facial images or\\ dactyloscopic data} \\
         \hline
         Non-personal data & \tabincell{l}{The definition is referred to as data other than personal data as\\ defined in Article 4, point (1), of Regulation (EU) 2016/679\\ (EU AI Act)} & \tabincell{l}{Locally produced\\ weather forecast,\\ energy consumption\\ prediction, demand\\ forecast of buildings} \\
         \hline
         Security & \tabincell{l}{Their is no universal, stand‑alone definition of security particularly for\\ AI and information system, to the best of our knowledge. Below we\\ provide the most relevant regulatory definitions closely related to security} & N/A \\
         \hline
         \tabincell{l}{Network and\\ information security} & \tabincell{l}{The ability of a network or an information system to resist, at a given\\ level of confidence, accidental events or unlawful or malicious actions\\ that compromise the availability, authenticity, integrity and\\ confidentiality of stored or transmitted data and the related services\\ offered by or accessible via those networks and systems\\ (Regulation (EU) No 526/2013)} & N/A \\
         \hline
         Cybersecurity & \tabincell{l}{Referred to Regulation (EU) 2019/881 (EU AI Act)\\ The activities necessary to protect network and information systems,\\ the users of such systems, and other persons affected by cyber threats\\ (Regulation (EU) 2019/881)} & N/A \\
         \hline
         \tabincell{l}{Information system} & \tabincell{l}{Computers and electronic communication networks, as well as electronic\\ data stored, processed, retrieved or transmitted by them for the\\ purposes of their operation, use, protection and maintenance\\ (Regulation (EC) No 460/2004)} & \tabincell{l}{Telecommunication\\ networks, banking\\ systems, industry\\ control systems.} \\
         \hline
         Systemic risk& \tabincell{l}{A risk that is specific to the high-impact capabilities of\\ general-purpose AI models, having a significant impact on the Union\\ market due to their reach, or due to actual or reasonably foreseeable\\ negative effects on public health, safety, public, security, fundamental\\ rights, or the society as a whole, that can be propagated at scale across\\ the value chain. (EU AI Act)}  & \tabincell{l}{AI tools that\\ massively produce\\ realistic deep‑fake\\ figures, news, or\\ coordinated\\ disinformation\\ campaigns on social\\ media.} \\
         \hline
    \end{tabular}
    }
\end{table}

We also notice that EU regulations refer a lot to information system when talking about security. We would like to note that the relationship between information system and AI system is fundamental. AI represents a specialized layer of logic and inference that resides within the broader of an information system, and AI  cannot exist without an information system to provide services. An AI system can be viewed as an application of an AI model integrated into an information system with computation power, data storage, and network resources. Further, the presence of AI, especially agentic AI, increases the security attack surface of the information systems. This is because agentic AI is no longer just a ``resident'' or ``tenant'' of the information system, but a ``user'' or ``operator'' that can interact with the information system autonomously, \textit{e.g.}, in executing code or modifying database. Therefore, an attack on the agentic AI becomes a direct security breach of the information system, in the formats network and information security or cybersecurity, or other new types of issues.

Information systems are traditionally audited through code reviews and network logs. However, because LLMs and GPAI are ``probabilistic", or black boxes, they introduce non-deterministic risks into otherwise deterministic information systems. Regulation now requires ``AI-specific" cybersecurity measures, such as input/output filtering and adversarial testing.

AI system is viewed as a high-level functional layer that operates within the broader context of an information system. While an information system provides the capabilities for operation, the AI System provides the capabilities for inference. As AI moves from passive generation (LLMs) to active orchestration (Agentic AI), the legal distinction between ``the tool" and ``the user" of the information system continues to blur, requiring a unified approach to digital governance.

\section{Regulatory provisions}\label{secc:provision}

In this section, we look into privacy and security provisions from the reviewed EU AI regulatory documents. We analyze and reflect the provisions pertinent to different types of AI systems, and check exactly the status of regulatory provisions targeting agentic AI.

\subsection{Privacy provisions}

We list the provisions towards privacy and personal data protection for AI systems and GPAI in Table~\ref{tab:privacy_non_agentic_AI}. Note that there is no provisions specifically targeting privacy for GAI and LLMs. From the table, we observe that the regulatory documents integrates the principles in privacy and personal data protection laid down in existing regulations in AI related regulatory documents to guide how AI should be used. We also notice that while there are sufficient privacy provisions targeting AI systems, those dedicated to specific types of AI are rare. AI providers and practitioner, regardless of what type of AI they are using, can follow the provisions for AI systems in they practice in general, so as to pursuit regulatory compliance. However, more specific provisions for different types of AI can help reduce regulatory uncertainties and promote AI literacy\footnote{``AI literacy'' means skills, knowledge and understanding that allow providers, deployers and affected persons, taking into account their respective rights and obligations in the context of this Regulation, to make an informed deployment of AI systems, as well as to gain awareness about the opportunities and risks of AI and possible harm it can cause, Article 3 (56) of Regulation (EU) 2024/1689 (AI Act).}, thus facilitating development in AI. 

\begin{table}[htbp]
    \centering
    \caption{Privacy and personal data protection provisions for different types of AI}\label{tab:privacy_non_agentic_AI}
    \resizebox{0.95\textwidth}{!}{
    \begin{tabular}{ll}
        \hline
        AI concepts & Privacy and personal data protection provisions (not a exhaustive list)\\
         \hline
        AI systems & \tabincell{l}{AI systems should make best efforts to respect general principles ... in line with the Charter of\\ Fundamental Rights of the European Union ..., including the protection of fundamental rights,\\ human agency and oversight, technical robustness and safety, \textbf{privacy and data governance}, ....\\ (Amendments adopted for the proposal on AI Act)\\ The indiscriminate and untargeted scraping of biometric data from social media or CCTV\\ footage to create or expand facial recognition databases add to the feeling of mass surveillance\\ and can lead to gross violations of fundamental rights, including \textbf{the right to privacy}. The use\\ of AI systems with this intended purpose should therefore be prohibited\\ (Amendments adopted for the proposal on AI Act)\\ Throughout the recruitment process ... of persons ..., such systems may perpetuate historical patterns\\ of discrimination, for example against women, certain age groups, ... . AI systems used to monitor\\ the performance and behaviour of these persons may also undermine the essence of their fundamental\\ rights to \textbf{data protection and privacy}. This Regulation applies without prejudice to Union and\\ Member State competences to provide for more specific rules for the use of AI-systems in the\\ employment context (Amendments adopted for the proposal on AI Act)\\ ... for the training, validation and testing of AI systems ... the European health data space will\\ facilitate non-discriminatory access to health data and the training of artificial intelligence algorithms\\ on those datasets, in a \textbf{privacy-preserving}, secure, timely, transparent and trustworthy manner\\ (Amendments adopted for the proposal on AI Act)\\ The right to \textbf{privacy and to protection of personal data} must be guaranteed throughout the entire\\ lifecycle of the AI system. ... data minimisation and data protection by design and by default ... are\\ essential when the processing of data involves significant risks to the fundamental rights of individuals.\\ Providers and users of AI systems should implement state-of-the-art technical and organisational\\ measures ... . Such measures should include not only anonymisation and encryption, but also the use\\ of increasingly available technology ... (Amendments adopted for the proposal on AI Act)\\ To the extent that it is strictly necessary for the purposes of ensuring negative bias detection and\\ correction in relation to the high-risk AI systems, the providers ... may exceptionally process \textbf{special} \\\textbf{categories of personal data} ... subject to appropriate safeguards ... including technical limitations\\ on the re-use and use of state-of-the-art security and privacy-preserving \\ (Amendments adopted for the proposal on AI Act)\\ Any processing of \textbf{biometric data} and other \textbf{personal data} involved in the use of AI systems for\\ biometric identification, other than in connection to the use of real-time remote biometric identification\\ systems in publicly accessible spaces for the purpose of law enforcement as regulated by this Regulation,\\ should continue to comply with all requirements resulting from Article 10 of Directive (EU) 2016/680\\ (EU AI Act)\\ ... suggests including and documenting \textbf{privacy} and data security measures ... in the development and\\ training of AI. The inclusion and documentation of robust \textbf{privacy} and data security measures, including\\ encryption, access controls, and regular audits, protect sensitive data from cyber threats\\ (Opinion on ehtical AI and access to supercomputing for start-ups)}\\
         \hline
        GPAI & \tabincell{l}{General-purpose AI models could pose systemic risks ... In particular, international approaches have so\\ far identified the need to pay attention to risks ... the facilitation of disinformation or harming \textbf{privacy}\\ with threats to democratic values and human rights (EU AI Act)\\ ... the AI Act allows providers of high-risk AI systems to exceptionally use \textbf{sensitive personal data}\\ – which is otherwise prohibited by the GDPR – for the purpose of bias detection and correction. This\\ facilitates effective AI training and testing. The possibility of relying on this legal basis should be\\ extended to providers of all AI systems and general-purpose AI models\\ (EC Commission Staff Working Document on Digital Ommibus on AI)}\\
         \hline
    \end{tabular}
    }
\end{table}

While the above shows analysis on privacy provisions for non agentic AI, we note that for now, there are no provisions dedicated to agentic AI. Therefore, it is currently not clear whether specific measures should be conducted for agentic AI when it comes to regulatory privacy in EU. Though it is the same situations that agentic AI practitioners and providers can refer to general privacy provisions for compliance purpose, the interpretation of generally rules can lead to obscures and uncertainties that hurdles their business.

\begin{table}[bthp]
    \centering
    \caption{Security provisions for different types of AI}\label{tab:security_non_agentic_AI}
    \resizebox{0.95\textwidth}{!}{
    \begin{tabular}{ll}
        \hline
        AI concepts & Security provisions (not a exhaustive list)\\
        \hline
        AI systems & \tabincell{l}{... AI Act will guarantee the use of trustworthy AI and ensure the transparency, safety and required\\ human oversight. In addition, complementary regulations ensuring \textbf{cybersecurity} and privacy are key\\ ... mitigating the risk of potential misuse ... particularly in contexts like biowarfare\\ (EC Communication on boosting startups and innovation in trustworthy AI)\\ High-risk AI systems should perform consistently throughout their lifecycle and meet an appropriate\\ level of accuracy, robustness and \textbf{cybersecurity} (EU AI Act)\\ To ensure a level of \textbf{cybersecurity} appropriate to the risks, suitable measures, such as \textbf{security}\\ controls, should therefore be taken by the providers of high-risk AI systems, also taking into account as\\ appropriate the underlying ICT infrastructure (EU AI Act)\\ ... the assessment of the \textbf{cybersecurity} risks, associated to a product with digital elements\\ classified as high-risk AI system ... should consider risks to the cyber resilience of an AI system as\\ regards attempts by unauthorised third parties to alter its use, behaviour or performance, including AI\\ specific vulnerabilities such as data poisoning or adversarial attacks ... (EU AI Act)}\\
        \hline
        GAI & \tabincell{l}{Generative AI can exponentially increase the capacity to learn and replicate patterns found in cyber\\ threats ... thereby assisting \textbf{cybersecurity} professionals ... generative AI can also be used by\\ cybercriminals to organise sophisticated cyber-attacks ... internal \textbf{security} actors will also need to be\\ well equipped to address the use of generative AI by cybercriminals\\ (EC Communication on boosting startups and innovation in trustworthy AI)} \\
        \hline
        GPAI & \tabincell{l}{The providers of general-purpose AI models presenting systemic risks should be subject ... to obligations\\ aimed at identifying and mitigating those risks and ensuring an adequate level of \textbf{cybersecurity}\\ protection, regardless of whether it is provided as a standalone model or embedded in an AI system or a\\ product (EU AI Act)\\ ... providers (of general-purpose AI models) should ensure an adequate level of \textbf{cybersecurity} protection for\\ the model and its physical infrastructure, if appropriate, along the entire model lifecycle. \textbf{Cybersecurity}\\ protection related to systemic risks associated with malicious use or attacks should duly consider accidental\\ model leakage, unauthorised releases, circumvention of safety measures, and defence against cyberattacks,\\ unauthorised access or model theft. That protection could be facilitated by securing model weights,\\ algorithms, servers, and data sets, such as through operational security measures for \textbf{information security},\\ specific \textbf{cybersecurity} policies, adequate technical and established solutions, and cyber and physical access\\ controls ... (EU AI Act)} \\
        \hline
    \end{tabular}
    }
\end{table}

This gap reflects the rapid technological development in AI and the need of swift follow-up by the regulation side, and the need of contextualized provisions for privacy, particularly for the emerging concept of agentic AI. Actually, there are learned experience for the contextualization of generally regulations in specific areas, \textit{e.g.}, when it comes to privacy and personal data protection in smart grids. The solutions is the regulatory efforts made by dedicated EU expert groups, where they draft regulatory documents to interpret and contextualize privacy in smart grid. Their regulatory documents, \textit{e.g.}, the Data Protection Impact Assessment (DPIA) template for smart grid\footnote{\url{https://energy.ec.europa.eu/document/download/eee93bb8-1bda-4bdc-ac64-7edd6d0e60bc_en?filename=dpia_for_publication_2018.pdf}, accessed 2026-01-15.}, are recognized by EU regulations, and combining with the general privacy regulations, it formulates full-fledged regulatory privacy and personal data protection in the smart grid. Taking this experience, we suggest dedicate efforts from the regulation side that can improve regulatory literacy and mitigate uncertainties.

\subsection{Security provisions}

We show security provisions for AI from the reviewed regulatory documents in Table~\ref{tab:security_non_agentic_AI}. Similar to the case of privacy, security is not well contextualized in different types of AI. For now, dedicated provisions are available for the concepts of AI system, GAI, and GPAI, while missing for LLMs and agentic AI. Even though, it is applicable that practitioners of LLMs and agentic AI take the security provisions for other types of AI in their practice, while subject to a regulatory uncertainty. 

Furthermore, we notice that the security provisions for AI are generally risk orientated. That is, those provisions either target high-risk AI system, or systemic risks potentially posed by GAI or GAPI. Therefore, we envision that the definition boundaries for such risks and systems will be critical, so it is with the interpretation and exemplification of the definitions that can facilitate intuitive and crucial understanding of AI practitioner in their business. We notice that while a general regulatory boundary for ``high-risk AI systems'' is available, it will be even more useful if a more granular illustration/interpretation/exemplification on different types of AI can be provided. 

\section{Conclusions}\label{sec:conclusion}

This work reviews EU AI regulatory documents that are published between 2024 and 2025. We check regulatory definitions related to privacy, security, and agentic AI, and distinct them from closely relevant concepts. We examine the regulatory provisions for privacy, security, and agentic AI, and analyze and reflect the status of the provisions for AI especially for the emerging concept of agentic AI. Our study reveals that, though applicable provisions exist, the contextualization of the provision for AI particularly for different types of AI remains in its early stage. We also show the connections between the provisions for AI and those for general information systems and indiscriminative areas, so as to bridge AI provisions with a broader context in regulation compliance. We envision that future efforts are needed in mitigating regulatory uncertainties, by differentiating, interpreting, and articulating privacy and security provisions for AI in a more granular manner. 

\section*{Acknowledgment}

This work was supported by the PriTEM project funded by UiO:Energy Convergence Environments.

\bibliographystyle{ieeetr}
\bibliography{references}

@article{radanliev2025artificial,
  title={Artificial intelligence: reflecting on the past and looking towards the next paradigm shift},
  author={Radanliev, Petar},
  journal={Journal of Experimental \& Theoretical Artificial Intelligence},
  volume={37},
  number={7},
  pages={1045--1062},
  year={2025},
  publisher={Taylor \& Francis}
}

@article{melguizo2026can,
  title={Can {AI} grow green? Evidence of a Kuznets curve among {AI}, renewable energies and emissions},
  author={Melguizo, Angel and Katz, Ra{\'u}l and Jung, Juan},
  journal={Energy Policy},
  volume={208},
  pages={114883},
  year={2026},
  publisher={Elsevier}
}

@article{kim2025understanding,
  title={Understanding user preferences in developing a mental healthcare {AI}  chatbot: a conjoint analysis approach},
  author={Kim, Mirae and Oh, Jaedong and Kim, Doha and Shin, Jungwoo and Lee, Daeho},
  journal={International Journal of Human--Computer Interaction},
  volume={41},
  number={8},
  pages={4813--4821},
  year={2025},
  publisher={Taylor \& Francis}
}

@article{vukovic2025ai,
  title={{AI} integration in financial services: a systematic review of trends and regulatory challenges},
  author={Vukovi{\'c}, Darko B and Dekpo-Adza, Senanu and Matovi{\'c}, Stefana},
  journal={Humanities and Social Sciences Communications},
  volume={12},
  number={1},
  pages={1--29},
  year={2025},
  publisher={Palgrave}
}

@article{fosso2026building,
  title={Building {AI}-enabled capabilities for improved environmental and manufacturing performance: evidence from the US and the UK},
  author={Fosso-Wamba, Samuel and Guthrie, Cameron and Queiroz, Maciel M and Oyedijo, Adegboyega},
  journal={International Journal of Production Research},
  volume={64},
  number={2},
  pages={545--564},
  year={2026},
  publisher={Taylor \& Francis}
}

@article{yan2025generative,
  title={Generative {AI}  for Intelligent Transportation Systems: Road Transportation Perspective},
  author={Yan, Huan and Li, Yong},
  journal={ACM Computing Surveys},
  year={2025},
  publisher={ACM New York, NY}
}

@article{spencer2025ai,
  title={{AI}, automation and the lightening of work},
  author={Spencer, David A},
  journal={AI \& society},
  volume={40},
  number={3},
  pages={1237--1247},
  year={2025},
  publisher={Springer}
}

@article{chen2025uncovering,
  title={Uncovering the results of {AI}  Chatbot use in the public sector: Evidence from US state governments},
  author={Chen, Tzuhao and Gasco-Hernandez, Mila},
  journal={Public Performance \& Management Review},
  volume={48},
  number={6},
  pages={1331--1356},
  year={2025},
  publisher={Taylor \& Francis}
}

@article{tan2025artificial,
  title={Artificial intelligence in teaching and teacher professional development: A systematic review},
  author={Tan, Xiao and Cheng, Gary and Ling, Man Ho},
  journal={Computers and Education: Artificial Intelligence},
  volume={8},
  pages={100355},
  year={2025},
  publisher={Elsevier}
}

@article{leo2026threat,
  title={From threat to trust: assessing security risks of agentic {AI}  systems},
  author={Leo, Martin and Tan, Freedy and Miao, Tianqi and Anand, Guru},
  journal={International Journal of Information Security},
  volume={25},
  number={1},
  pages={23},
  year={2026},
  publisher={Springer}
}

@article{hasselwander2026toward,
  title={Toward agentic {AI}: User acceptance of a deeply personalized {AI}  super assistant ({AISA})},
  author={Hasselwander, Marc and Sunio, Varsolo and Lah, Oliver and Mogaji, Emmanuel},
  journal={Journal of Retailing and Consumer Services},
  volume={89},
  pages={104620},
  year={2026},
  publisher={Elsevier}
}

@book{biswas2025building,
  title={Building Agentic {AI}  Systems: Create intelligent, autonomous {AI} agents that can reason, plan, and adapt},
  author={Biswas, Anjanava and Talukdar, Wrick},
  year={2025},
  publisher={Packt Publishing Ltd}
}

@article{hosseini2025role,
  title={The role of agentic {AI} in shaping a smart future: A systematic review},
  author={Hosseini, Soodeh and Seilani, Hossein},
  journal={Array},
  pages={100399},
  year={2025},
  publisher={Elsevier}
}

@article{deng2025ai,
  title={{AI} agents under threat: A survey of key security challenges and future pathways},
  author={Deng, Zehang and Guo, Yongjian and Han, Changzhou and Ma, Wanlun and Xiong, Junwu and Wen, Sheng and Xiang, Yang},
  journal={ACM Computing Surveys},
  volume={57},
  number={7},
  pages={1--36},
  year={2025},
  publisher={ACM New York, NY}
}

@article{ferrag2025prompt,
  title={From prompt injections to protocol exploits: Threats in {LLM}-powered {AI} agents workflows},
  author={Ferrag, Mohamed Amine and Tihanyi, Norbert and Hamouda, Djallel and Maglaras, Leandros and Lakas, Abderrahmane and Debbah, Merouane},
  journal={ICT Express},
  year={2025},
  publisher={Elsevier}
}

@incollection{huang2025securing,
  title={Securing Multi-Modal Agentic {AI} Systems},
  author={Huang, Ken and Hughes, Chris},
  booktitle={Securing {AI} Agents: Foundations, Frameworks, and Real-World Deployment},
  pages={253--285},
  year={2025},
  publisher={Springer}
}

@incollection{huang2025commercial,
  title={The Commercial Landscape of Agentic {AI} Security},
  author={Huang, Ken and Hughes, Chris},
  booktitle={Securing {AI} Agents: Foundations, Frameworks, and Real-World Deployment},
  pages={347--373},
  year={2025},
  publisher={Springer}
}

@article{jannelli2025agentic,
  title={Agentic {LLM}s in the supply chain: towards autonomous multi-agent consensus-seeking},
  author={Jannelli, Valeria and Schoepf, Stefan and Bickel, Matthias and Netland, Torbj{\o}rn and Brintrup, Alexandra},
  journal={International Journal of Production Research},
  pages={1--31},
  year={2025},
  publisher={Taylor \& Francis}
}

@article{esen2025risks,
  title={The Risks of Agentic {AI}: The Curse of Autonomy},
  author={ESEN, Fatih Sinan},
  journal={The Age of Generative Artificial Intelligence},
  pages={156},
  year={2025},
  publisher={{\.I}zmir Akademi Derne{\u{g}}i}
}

@article{ray2026comprehensive,
  title={A comprehensive introspection on {AI} risks: taxonomy, challenges, and future directions},
  author={Ray, Partha Pratim},
  journal={Iran Journal of Computer Science},
  volume={9},
  number={1},
  pages={18},
  year={2026},
  publisher={Springer}
}

@inproceedings{zhang2025towards,
  title={Towards Aligning Personalized {AI} Agents with Users' Privacy Preference},
  author={Zhang, Shuning and Ma, Ying and Chen, Jingruo and Li, Simin and Yi, Xin and Li, Hewu},
  booktitle={Proceedings of the 2025 Workshop on Human-Centered {AI} Privacy and Security},
  pages={33--42},
  year={2025}
}

@article{dwivedi2025agentic,
  title={Agentic {AI} Systems: What It Is and Isn't},
  author={Dwivedi, Yogesh K and Helal, Mohamed YI and Elgendy, Ibrahim A and Alahmad, Rasha and Walton, Paul and Suh, Ayoung and Singh, Vinay and Jeon, Il},
  journal={Global Business and Organizational Excellence},
  year={2025},
  publisher={Wiley Online Library}
}

@misc{hughes2025ai,
  title={AI Agents and Agentic Systems: Redefining Global it Management},
  author={Hughes, Laurie and Dwivedi, Yogesh K and Li, Keyao and Appanderanda, Mandanna and Al-Bashrawi, Mousa Ahmad and Chae, Inyoung},
  journal={Journal of Global Information Technology Management},
  pages={1--11},
  year={2025},
  publisher={Taylor \& Francis}
}

@inproceedings{nolte2025robustness,
  title={Robustness and cybersecurity in the {EU} artificial intelligence act},
  author={Nolte, Henrik and Rateike, Miriam and Finck, Mich{\`e}le},
  booktitle={Proceedings of the 2025 ACM Conference on Fairness, Accountability, and Transparency},
  pages={283--295},
  year={2025}
}

@article{beltran2025ai,
  title={{AI algorithms under scrutiny: GDPR, DSA, AI Act and CRA as pillars for algorithmic security and privacy in the European Union}},
  author={Beltr{\'a}n, Marta},
  journal={Computers \& Security},
  pages={104628},
  year={2025},
  publisher={Elsevier}
}

@article{kianpour2025digital,
  title={Digital sovereignty in practice: analyzing the {EU}’s {NIS}2 directive},
  author={Kianpour, Mazaher and Earls Davis, Peter Alexander and Windekilde, Iwona Maria},
  journal={International Journal of Information Security},
  volume={24},
  number={4},
  pages={1--11},
  year={2025},
  publisher={Springer}
}

@incollection{mueck2025introduction,
  title={Introduction to the European Cyber Resilience Act},
  author={Mueck, Markus and Roberts, Taylor and Du Boisp{\'e}an, St{\'e}phane and Gaie, Christophe},
  booktitle={European Digital Regulations},
  pages={91--110},
  year={2025},
  publisher={Springer}
}

@article{hohmann2025reflections,
  title={Reflections on the data protection compliance of {AI}  systems under the {EU} {AI}  Act},
  author={Hohmann, Bal{\'a}zs and Koll{\'a}r, Gerg{\H{o}}},
  journal={Cogent Social Sciences},
  volume={11},
  number={1},
  pages={2560654},
  year={2025},
  publisher={Taylor \& Francis}
}

@article{calvano2026building,
  title={Building Symbiotic Artificial Intelligence: Reviewing the {AI}  Act for a Human-Centred, Principle-Based Framework},
  author={Calvano, Miriana and Curci, Antonio and Desolda, Giuseppe and Esposito, Andrea and Lanzilotti, Rosa and Piccinno, Antonio},
  journal={Minds and Machines},
  volume={36},
  number={1},
  pages={1},
  year={2026},
  publisher={Springer}
}

@article{nizza2026ais,
  title={What do {AI}s think About the {AI} {A}ct? An Experimental Analysis of the {EU} Approach on Artificial Intelligence},
  author={Nizza, Umberto},
  journal={EUROPEAN BUSINESS LAW REVIEW},
  volume={36},
  number={2},
  year={2026}
}

@article{sharma2026generative,
  title={Generative {AI} and digital twins for sustainable last-mile logistics: Enabling green operations and electric vehicle integration},
  author={Sharma, Shashikant Nishant},
  journal={Accelerating logistics through generative AI, digital twins, and autonomous operations},
  pages={183--216},
  year={2026},
  publisher={IGI Global Scientific Publishing}
}

@article{zhang2025data,
  title={Data Sharing, Privacy and Security Considerations in the Energy Sector: A Review from Technical Landscape to Regulatory Specifications},
  author={Zhang, Shiliang and Maharjan, Sabita and Bygrave, Lee Andrew and Yu, Shui},
  journal={arXiv preprint arXiv:2503.03539},
  year={2025}
}

\end{document}